\documentclass{aa} 
\usepackage{graphicx} 
\usepackage{natbib}
\usepackage{aalongtable}
\bibpunct{(}{)}{;}{a}{}{,}
\usepackage{txfonts}

\begin{document}
\title{Size distribution of circumstellar disks in the Trapezium
  cluster}

\author{S\'{\i}lvia M. Vicente\inst{1,2} \and Jo\~ao Alves\inst{1}}

\offprints{S. Vicente} \mail{svicente@eso.org}
\institute{$^1$European Southern Observatory, Karl-Schwarzschild
  Stra\ss e 2, D-85748 Garching bei M\"unchen, Germany\\ $^2$Faculdade
  de Ci\^encias da Universidade de Lisboa, Campo Grande, Portugal\\
  e--mail: svicente@eso.org, jalves@eso.org}

\date{Received / Accepted 3 June 2005}

\titlerunning{Size Distribution of Circumstellar Disks}
\authorrunning{S. Vicente \& J. Alves}

\abstract{In this paper we present results on the size distribution of
  circumstellar disks in the  Trapezium cluster as measured from
  HST/WFPC2 data.  Direct diameter measurements of a sample of 135
  bright proplyds and 14 silhouettes disks suggest that there is a
  single population of disks well characterized by a power-law
  distribution with an exponent of $-$1.9 $\pm$ 0.3 between disk
  diameters 100--400 AU. For the stellar mass sampled (from late G to
  late M stars) we find no obvious correlation between disk diameter
  and stellar mass. We also find that there is no obvious correlation
  between disk diameter and the projected distance to the ionizing Trapezium OB
  stars. We estimate that about 40\% of the disks in the Trapezium
  have radius larger than 50 AU. We suggest that the origin of the
  Solar system's (Kuiper belt) outer edge is likely to be due to the
  star formation environment and disk destruction processes
  (photoevaporation, collisions) present in the stellar cluster on
  which the Sun was probably formed. Finally, we identified a
  previously unknown proplyd and named it 266-557, following
  convention.}

 \maketitle


%

\section{Introduction}

Understanding the formation of pla\-ne\-ts from protoplanetary disks
surrounding young stars is an important goal of mo\-dern day
astrophysics. Although there is e\-vi\-dence for most stars in our
Galaxy to have formed in dense photo-evaporating clusters much like the
 Trapezium cluster (e.g., Lada \& Lada 2003), it is not yet clear that
the circumstellar disks once associated with these stars survived the
harsh conditions imposed by massive star formation, and matured into
planetary systems.  The  {Trapezium cluster} in the Orion star forming
region ( M42,  NGC 1976) is a unique laboratory to tackle this problem,
mainly because it is the closest (d$=$450 pc, e.g., Muench et al.
2002) massive star forming region.  It is also very young ($\sim$ 1
Myr) (Hillenbrand 1997; Hillenbrand \& Carpenter 2000), and harbors
nearly 2\,000 YSOs within a few parsecs (Hillenbrand \& Hartmann
1998), the cluster core alone containing over 100 YSOs stars within
0.1 pc of the star $\theta^{1}$ Ori C.

Solar system-sized circumstellar disks in the  Orion Nebula were first
inferred from radio observations of compact ionized regions
surroun\-ding young low-mass stars (Churchwell et al.  1987). The
Hubble Space Telescope (HST) subsequently provided the most compelling
evidence for disks in the spectacular images of extended circumstellar
material surrounding young low-mass stars in the core of the Orion
Nebula.  Over half of the 300 YSO observed in the HST images were
classified as proplyds (PROtoPLanetarY DiskS) (O'Dell et al. 1993,
O'Dell \& Wen 1994; O'Dell \& Wong 1996; McCaughrean \& O'Dell 1996;
Bally et al. 1998a; Bally et al. 2000), flattened circumstellar clouds
of dust and gas surrounding young stars rendered visible through
inclusion in or proximity to an H {\small II} region. The in\-ten\-se UV
radiation fields of the massive OB stars heat the disk surface, drive
mass-loss and produce bright io\-ni\-za\-ti\-on fronts due to the
emission-line radiation arising from the outer parts of the proplyds.
Most of them are comet-shaped ionized envelopes pointing directly away
from the brightest OB stars and are believed to contain evaporating
circumstellar disks (McCaughrean et al. 1998; Johnstone et al. 1998).
The principal sources of ionizing radiation in the region are the O6,
45 M$_\odot$ star $\theta^{1}$ Orionis C, the brightest of the
Trapezium compact group of 4 high-mass OB stars, and the O9.5,
25M$_\odot$ star $\theta^{2}$ Orionis A located several arcminutes to
the south at a distance of 0.3 pc from $\theta^{1}$ Ori C. Dusty disks
are seen either as dark silhouettes against the bright background
nebular light -- the pure silhouettes -- (McCaughrean \& O'Dell 1996;
Bally et al. 2000) or embedded in the light from their own ionization
fronts -- the embedded silhouettes (Bally et al. 2000).  Proplyds
display a variety of forms. The ones closer to $\theta^{1}$ Ori C have
bright cusps in optical emission lines (H$_\alpha$,\- [O {\small III}],\- [N
{\small II}]) facing $\theta^{1}$ Ori C and ``tails'' extending from the ends
of the cusps. The farthest have curved and close boundaries that
include the entire circumstellar cloud.   In the fields of the inner
portion of the  Orion Nebula -- a region that contains more than 300
YSO of the nearly 2000 members of the extended  Trapezium cluster --
imaged with the HST, 161 proplyds were identified, of which 15 are
pure silhouettes and 146 are bright droplets surrounded by ionization
fronts (IF).

O'Dell \cite{odell01a} examined WFPC2/HST parallel planned observations in the
outer portions of the  Orion Nebula and near the center of its
companion H {\small II} region M43 located several arc\-mi\-nutes north of
the Trapezium and powered by a single B0 spectral type star,  NU
Ori. His discovery of three new bright proplyds (093-822; 307-1807;
 332-1605)  and one pure sillhouette proplyd (321-602) is an important
proof that additional young stars, disks and jets remain to be
revealed in the outer parts of the  Orion Nebula.  Subsequently, Smith
et al. \cite{smith05}, as a result of an H$_\alpha$ survey with the HST Wide
Field Camera of the Advanced Camera for Surveys (ACS/WFC), discovered
10 new silhouette disks in the outskirts of the  Orion Nebula and in its
neighboring region  M43: 6 pure silhouettes -- 053-717, 110-3035,
141-1952, 280-1720, 347-1535, 216-0939 -- and 4 embedded silhouettes -- 
132-042, 124-132, 253-1536, 181-826 (this one discussed in Bally et
al. 2005). These new disks exhibit extended emission from bipolar
reflection nebulae and microjets due to a fainter background in the
less intense radiation fields regions far from the Trapezium core.
Up to date pu\-bli\-cations, 171 proplyds were imaged with
the \textit{Hubble Space Telescope} (HST) in the  Orion Nebula and  M43: 
150 bright cusps and 21 pure silhouettes.

In this paper we analyze available HST-WFPC2 images to measure the
diameters of 149 proplyds (14 are pure silhouettes and 135 are bright
proplyds surrounded by ionization fronts) and derive the basic
statistics of this population. These statistics can provide important
constraints to models of protoplanetary disk evolution and planet
formation in young stellar clusters.  The observations are described
in \S2 and the results presented in \S3.  The results are discussed in
\S4 and the main conclusions are summarized in \S5.

\section{Observations}

To perform the diameter measurements we selected from the available
\textit {Hubble Space Telescope} (HST) images of the  Orion Nebula
(Bally et al. 2000)\footnote{\scriptsize The HST mosaicked ``master''
  images of the  Orion Nebula are accessible at
  http://casa.colorado.edu/$\sim$bally/HST/HST/master/.}  the one that
provided the best contrast between the disk and the background nebula
light. The available images are mosaic images from data acquired
through filters F673N, F658N, F656N, F631N, F547N and F502N, under the
general observer (GO) programs, GO 5085 (O'Dell \& Wong 1996) and GO
5469 (Bally et al. 1998a; O'Dell 1998).  After quantitative
comparisons we selected the H$_\alpha$ image since it provided the
best contrast and had the best signal-to-noise. 
The selected  H$_\alpha$ image covers an area in the sky of
7.8 $\times$ 8.3 arcmin$^{2}$ and has a resolution of about
0.$^{\prime\prime}$1.  World Coordinate System (WCS) coordinates were
inserted using IRAF\footnote{\scriptsize IRAF is distributed by the
  National Optical Astronomy Observatory, which is o\-pe\-ra\-ted by the
  Association of Universities for Research in Astronomy, Inc., under
  cooperative agreement with the National Science Foundation.} scripts
and the images were searched for circumstellar disks seen only in
silhouette, bright proplyds with dark disks, and bright proplyds
without visible disks.  The names of the objects discussed in this
paper use the coordinate-based system introduced by O'Dell \& Wen
\cite{odell94}, where the numbers are truncated positions (in the equinox
J2000.0 FK5 reference system) rounded off to 0$^s$.1 of right
ascension and 1$^{\prime\prime}$ in declination.
The first three coordinates correspond to the right ascension
(J2000.0) given to 0$^s$.1, with 5$^{h}$35$^{m}$ subtracted and the
decimal indicator dropped.  The last three digits correspond to the
declination (J2000.0) with 5$^\circ$20$^{\prime}$ added.  For example,
131-046 has a right ascension of 5$^{h}$35$^{m}$13$^s$.1 and a
declination of --5$^\circ$20$^{\prime}$46$^{\prime\prime}$. When the
object lies north of --5$^\circ$20$^{\prime}$, a four digit
declination indicator is used, like in 121-1925.

For the image analyzed in this paper the typical errors in the
positions are about 0.5 pixels or 0.$^{\prime\prime}$05.  For a
distance of 450 pc to the  Orion Nebula, the physical resolution of the
HST image is 0.$^{\prime\prime}$15 or $\sim$ 67.5 AU.  ($\sim$ 1.5
pixel/FWHM; 1 pixel corresponds to $\sim$ 0$^{\prime\prime}$.1 or 45
AU).  Bally et al. \cite{bally00} analyzed dithered images obtained with the
WFPC2 in the HST Cycle 6 under program GO 6603. The dithered images
were combined into final images using the drizzling technique
resulting in individual wide-field frames with 20\% better resolution
than the non-drizzled image analyzed in this paper. This has an impact
on the measurements of the proplyds' sizes, i.e., the diameters measured in
this paper are, on average, 20\% greater than the diameters listed in
Bally et al. \cite{bally00}.

\section{Results}
\subsection{The sample}
 
We assume circumstellar disks to be essentially circular structures
(typical disk thickness, $\sim 10$\% of diameter, is negligible when
compared to the precision of our measurements, see section 3.2). Projections
of a circular disk at any random orientation always corresponds to an
ellipse, exception given to the perfect edge-on case. The diameter of
the major axis of such an ellipse is always equal to the diameter of
the circular disk seen in projection.  Hence, a direct measurement of
the major axis of an observed ellipse is a direct measurement of the
disk diameter.

The initial sample consisted of 161 sources (see Table 3, available at the CDS), all
listed in previous papers (O'Dell et al. 1993, O'Dell \& Wen
1994; O'Dell \& Wong 1996, McCaughrean \& O'Dell 1996; Bally et al.
2000) but only 144 of them were identified in our analysis (14 were
not found, 2 were not included and 1 was out of the image). We found
an object which was clearly proplyd-like but its coordinates did not
correspond to any object in previous catalogs. We named it 266-557
using O'Dell's coordinate-based system and added it to the sample. For
the 14 proplyds not found in our analysis, 2 are silhouettes
(132-1832; 172-028) and 12 are bright proplyds and they are, either
indistinguishable from the background (064-705; 153-1902; 158-425;
166-519; 171-315; 174-400; 179-536; 205-421; 208-122; 215-317;
221-433), or very close to a bright star (163-322). We suggest that
some of their coordinates can be possibly wrong.  The proplyds
168-326NW and 168-326SE, listed as two diffe\-rent objects in previous
papers, were not resolved in the i\-ma\-ge as two objects, so they were
not included in the sample.  From the 15 silhouettes discovered until
and listed in the paper Bally et al. \cite{bally00}, the silhouette 294-606
appears out of the field of view in the H$_\alpha$ mosaic image we
used and, therefore, was not considered in the sample.  The proplyds
132-1832, 172-028, 171-315 and 205-421 were measured directly from
their pu\-bli\-shed images in the papers O'Dell \& Wen \cite{odell94} and Bally et
al. \cite{bally00}, resulting in a final sample of 149 proplyds where 14 are
pure silhouettes and 135 are bright proplyds. Table 3, available at the CDS, lists the
observed properties of all the 162 externally illuminated YSOs
considered in this paper, measured coordinates (RA and DEC), disk and
ionization front diameters in pixel and AU (measured in H$_\alpha$
emission), projected distances to $\theta^{1}$ Ori C and comments. See
discussion on the procedure followed in the next section.

O'Dell \cite{odell01a} and Smith at al. \cite{smith05} proplyds were not
included in our sample because they lie farther north from the region
covered by our H$_\alpha$ image.

\subsection{Procedure}

The measuring procedure can be summarized as the following: for the
dark silhouettes the diameter in pixel was considered to be the
diameter of an ellipse fitted to the $\sim$ 10\% contour below the
background intensity.  Because of the highly variable background in
the  Orion Nebula, the considered value for the intensity, is an
average (Imax + Imin)/2, of the immediate background surrounding the
disk.  For the bright proplyds, within the variety of dimensions that
can be measured, the most useful and practical is the ``chord
diameter'', meaning the distance between the tips of the cusps of the
proplyds as described by O'Dell \cite{odell98}.  This can be determined more
accurately than any other dimension, such as the size along the
direction from the ionizing star to the cusp, and should not vary with
different spatial orientations. But, the uncertainty of the chord
diameter is primarily determined by the judgment involved in where to
define the cusp boundary and amounts 1 to 2 pixels.  To overcome this
problem, we defined the cusp boundary as a circle fitted to the
contour of plus $\sim$ 30\% the average background intensity. The
``chord diameter'' is then the diameter of this circle with an
uncertainty of 0.5 pixel or 22.5 AU, the error as\-so\-ciated with the
measurement only.  Unfortunately, because of the highly variable
background intensity, the contrast between proplyds/silhouettes and
the background is sometimes low. In our sample of 135 bright proplyds,
37 objects, or 27\% of the total sample, were not bright enough so
they were fitted to a contour of +10\% the average background
intensity. At the same time, although clearly detected, 4 silhouettes
were not dark enough and were measured to plus a few percent.

\subsection{Do the proplyd diameters vary with distance from 
$\theta^{1}$ Orionis C ?}

The observational establishment of the proplyds as a well defined
class of objects has led to several attempts to theoretically model
them (Henney et al. 1996; Johnstone et al. 1998; Henney \& Arthur
1998; St\"orzer \& Hollenbach 1999; Richling \& Yorke 2000).  In this
section we will introduce the most important aspects of proplyd
photoevaporation models to stress that a correlation between proplyd
diameter and distance from the ionizing source is expected in theory.

The diameter of the hydrogen ionization fronts (bright proplyds) is a
complex function dominated by the incident UV photon flux or true
distance to the illuminating source and the disk radius.  The
appearance of the proplyds can be explained by the interaction of EUV
-- extreme-ultraviolet (Lyman continuum; \textit{h$\nu$}  $>$ 13.6 eV)
and FUV -- far-ultraviolet (6 eV $<$ \textit{h$\nu$} $<$ 13.6 eV)
radiation from an external source with a circumstellar disk.  The  FUV
photons are absorbed mainly by dust and penetrate much deeper
into the disk than the EUV photons. At re\-la\-ti\-ve high column densities
they dissociate molecules, heat the material impenetrable to the EUV
photons and initiate a neutral flow away from the disk. This
``photon-dominated region'' or PDR is encased within an hydrogen IF
where the outflowing PDR material can be ionized by the EUV photons.
Depending on the ratio I(UV)= FUV/EUV, the IF can stand off at a
considerable distance from the disk surface, appea\-ring as a bright
round head in the direction of the illuminating source with an
extended ``cometary'' tail of lower emissivity in the far side of the
disk.
Johnstone et al. \cite{johnstone98}, presented ana\-ly\-ti\-cal and
 numerical models of the structure of the neutral flows.  Depending on
 the disk radius, the I(UV)= FUV/EUV ratio and the column of neutral
 gas in the PDR, the neutral flow can be either FUV-dominated or
 EUV-dominated. For a given disk outer radius, as the distance to the
 ionization source increases, the EUV flux declines, I(UV) increases
 and the radius of the IF is expected to increase. The most rapid disk
 erosion occurs for large systems close to the Trapezium.
 St\"orzer \& Hollenbach \cite{storzer99} improved the previous Johnstone et al.
  \cite{johnstone98} model in\-clu\-ding the results of both equilibrium and
 non-equilibrium photo-dissociation region (PDR) codes to calculate
 the column den\-si\-ty and temperature inside the PDR. They determined
 under which circumstances FUV-dominated flows are possible and
 explained the observed IF diameters of the Orion proplyds by the
 FUV-dominated flows. These FUV-dominated flows are extended with IF
 radius r$_{IF}$ $\varpropto$ d$_{\theta^{1} Ori C}^{2/3}$ and
 typically, r$_{IF}$ $\gtrsim$ 2r$_d$, for a distance of 0.01--0.3 pc
 from $\theta^{1}$ Ori C. The mass-loss rate is proportional to the
 disk radius and not so dependent on the distance to $\theta^{1}$ Ori
 C.  Outside this region, EUV photons dominate the photoevaporation,
 the IF is closer to the disk surface, r$_{IF}$ $\lesssim$ 2r$_d$, and
 the mass-loss rate is proportional to d$_{\theta^{1} Ori C}^{-1}$ and
 r$_d^{3/2}$.

\textit {Is there any \textbf{observed} systematic variation of the proplyds diameter
with the distance to the main ionizing stars?}

%
\begin{figure}[t]
  \resizebox{9cm}{!}{\includegraphics[angle=0]{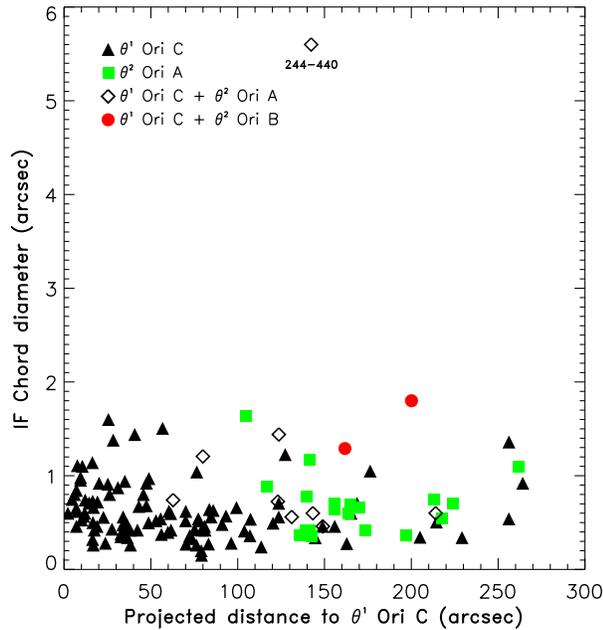}}
  \caption[]{Ionization front (IF) \textit {``chord diameter''} 
    (diameter of a circle fitted to the contour of plus $\sim$ 30\%
    the average background intensity), as a function of the projected
    distance to the main ionizing star in the  {Trapezium cluster},
    $\theta^{1}$ Ori C. The different symbols represent the proplyds
    that are ionized primarily by the stars indicated in the plot
    legend. For example, the largest bright proplyd 244-440 is being
    ionized by both $\theta^{1}$ Ori C and $\theta^{2}$ Ori A. There
    are 2 proplyds (252-457; 282-458) being ionized by both $\theta^{2}$
    Ori B, the bright star in the east direction from $\theta^{2}$ Ori
    A, and $\theta^{1}$ Ori C.}
  \label{fig:diameterIF}
\end{figure}
%

\begin{figure}[t]
  \resizebox{9cm}{!}{\includegraphics[angle=0]{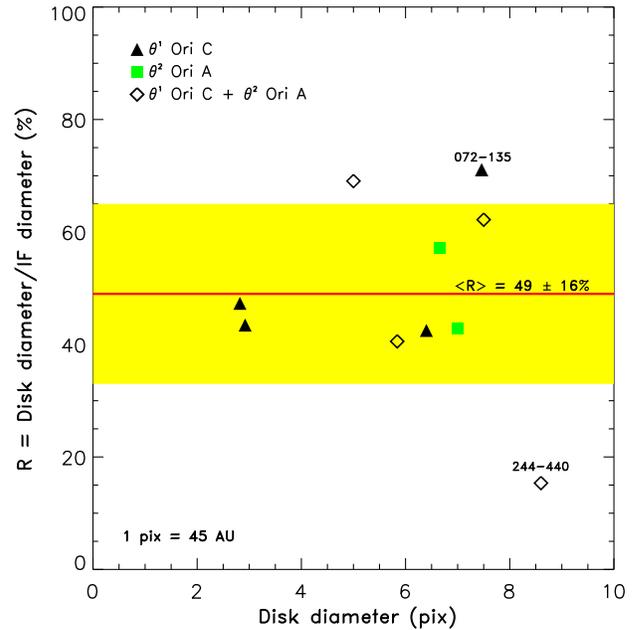}}
  \caption[]{Diameter ratio, R = disk diameter/ IF diameter, as a 
    function of the disk diameter in pixels. 
    The gray(yellow) band represents the average and dispersion of the sample
    (49 $\pm$ 16\%).}
  \label{fig:diameter_ratio}
\end{figure}

%

Figure 1 presents the IF chord diameter as a function of the projected
distance to $\theta^{1}$ Ori C for the 135 bright proplyds sample. The
different symbols indicate the main ionizing stars and, for each
proplyd, they were determined by the relative orientation of the
bright cusps and tails. For the not so clear cases, we drew a line in
our image connecting the ionizing stars and the center of the cusp and
compared them with the proplyd's tail-cusp direction.  The result of
this procedure was that there are 105 proplyds primarily ionized by
$\theta^{1}$ Ori C, 19 by $\theta^{2}$ Ori A, 9 by both $\theta^{1}$
Ori C and $\theta^{2}$ Ori A (like 244-440) and 2 io\-ni\-zed by
$\theta^{1}$ Ori C and $\theta^{2}$ Ori B (252-457 and 282-458).
Fig.1 clearly shows no obvious correlation between the ``chord
dia\-me\-ter'' and the projected distance to the ionizing Trapezium OB
stars. Assuming a random distribution of disk sizes across the
cluster, this result hints at poor correlation between IF size and 
distance, unlike theoretical predictions.

\subsection{Deriving  disk diameters from the IF chord diameters}

It is not trivial to infer disk diameters from the bright ionization
front diameters. Nevertheless, empirically, one can investigate if
there is a typical diameter ratio R (R = embedded disk
diameter/ionization front diameter) since we have in the sample bright
proplyds with clearly resolved embedded disks. We identified 10 such
cases (072-135; 141-520; 143-552; 163-222; 176-543; 181-247; 182-413;
197-427; 206-446; 244-440), at different projected distances from the
ionizing sources, and measured both the diameter of the ionization
front cusp and the diameter of the disk, as described in section 3.2.  Figure 2
represents the R values for the 10 silhouettes with embedded disks as
a function of the disk diameters.  The different symbols are related
to the different main ionizing stars. We find the average diameter
ratio for these 10 sources to be $<$R$>$ = 49 $\pm$ 16\%, a value
essentially identical to theoretical expectation for a distance of
0.01--0.3 pc from $\theta^{1}$ Ori C (St\"orzer \& Hollenbach 1999).
The observational parameters for these 10 proplyds are listed in Table
1.

\begin{table}
\caption[c]{Properties of the 10 proplyds with embedded silhouette disks.}
\label{table:1}
\tiny
\centering
\begin{tabular}{@{} lrrcccrcrc @{}}
\hline\hline
\noalign{\smallskip}
Name & \multicolumn{2}{r}{IF Diameter$^{\mathrm{a}}$} &\multicolumn{2}{c}{Disk Diameter$^{\mathrm{a}}$} & R & \multicolumn{2}{r}{d$_{proj} \theta^{1}$ OriC} & \multicolumn{2}{r}{d$_{proj} \theta^{2}$ OriA} \\ 
\noalign{\smallskip} 
& (pix) & (AU) & (pix) & (AU) & \% & ($\prime\prime$) & (pc) & ($\prime\prime$) & (pc) \\
\noalign{\smallskip}
\hline
\noalign{\smallskip}
072-135 & 10.5 (30\%) & 472 & 7.5 & 336 & 71 & 176.3 & 0.38 & 311.3 & 0.68 \\
141-520 & 7.2 (10\%) & 326 & 5.0 & 225 & 69 & 123.1 & 0.27 & 134.1 & 0.29 \\
143-522 & 14.4 (10\%) & 648 & 5.8 & 236& 40 & 123.7 & 0.27 & 130.4 & 0.28 \\
163-222 & 6.0 (10\%) & 268 & 2.8 & 127 & 47 & 61.4 & 0.13 & 185.2 & 0.40 \\
176-543 & 11.7 (10\%) & 525 & 6.7 & 300 & 57 & 141.3 & 0.31 & 91.8 & 0.20\\
181-247 & 6.7 (10\%) & 302 & 2.9 & 131 & 43 & 43.1 & 0.09 & 149.5 & 0.33 \\
182-413 & 15.1 (30\%) & 678 & 6.4 & 288 & 42 & 56.7 & 0.12 & 83.4 & 0.18 \\
197-427 & 12.1 (30\%) & 543 & 7.5 & 337 & 62 & 79.9 & 0.17 & 57.7 & 0.13 \\
206-446 & 16.3 (30\%) & 734 & 7.0 & 315 & 43 & 104.6 & 0.23 & 35.9 & 0.08 \\
244-440 & 56.0 (30\%) & 2520 & 8.6 & 387 & 15 & 142.3 & 0.31 & 29.2 & 0.06 \\
\noalign{\smallskip}
\hline
\end{tabular}
\begin{list} {}{}
\item[$^{\mathrm{a}}$] The disk and IF diameters were determined directly from the data and do not account for finite size of the \textit{HST} PSF.
\end{list}
\end{table}
Figure 3 is a spatial distribution diagram of the 10 proplyds with
embedded silhouette disks. The black filled circles re\-pre\-sent the
silhouette disks and the colored, unfilled disks represent the
ionization fronts. Green is for the proplyds ionized primarily by
$\theta^{1}$ Ori C (072-135; 163-222; 181-247; 182-413), blue for the
proplyds ionized by $\theta^{2}$ Ori A (176-543; 206-446) and red for
proplyds ionized by both stars (141-520; 143-522; 197-427; 244-440).
The circle diameters are proportional to the disk and IF diameters.
\begin{figure}[t]

  \resizebox{10cm}{!}{\includegraphics[angle=0]{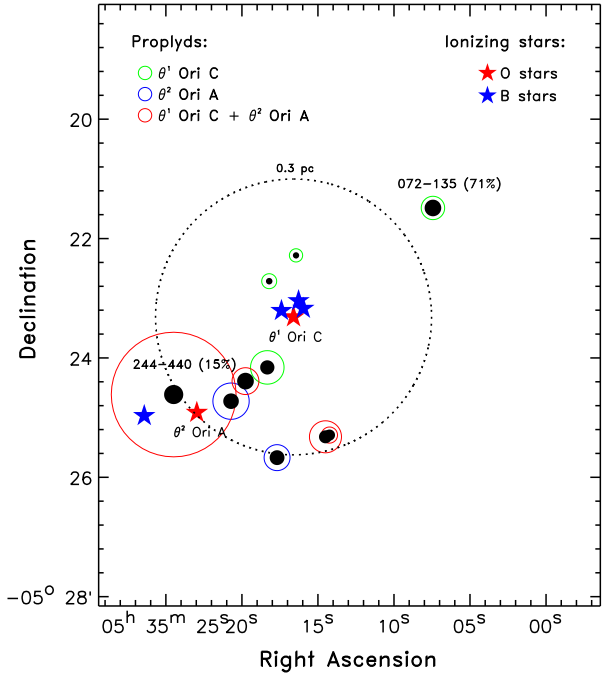}}
  \caption[]{Spatial distribution of the 10 proplyds with
    embedded silhou\-et\-te disks. The black filled circles represent the
    silhouette disks and the colored, open circles represent the
    ionization fronts. Green is for the proplyds ionized primarily by
    $\theta^{1}$ Ori C (072-135; 163-222; 181-247; 182-413), blue for
    the proplyds ionized by $\theta^{2}$ Ori A 
    (176-543; 206-446) and red for proplyds ionized by both stars
    (141-520; 143-522; 197-427; 244-440).  The circle diameters are proportional to the
    true disk and IF diameters. }
  \label{fig:spatial_10}
\end{figure}


There is a fairly large dispersion in the ratio R = disk dia\-me\-ter/IF
diameter, as showed by Fig. 2, but it is not clear that R is
correlated with the projected distance, or true distance, to
$\theta^{1}$ Ori C (see Fig.3).
For approximately the same proplyd disk diameter there are examples of
IFs very close to the disk and IFs that stand at a considerable
distances from the disk.  There is also no apparent increase of the
IF's diameters with the projected distance to $\theta^{1}$ Ori C
(Fig.1). The proplyds 244-440 (R = 15\%) and 072-135 (R = 71\%)
illustrate the two opposite ``extreme'' cases.  244-440 is localized
at a distance of 142.$^{\prime\prime}$3 from $\theta^{1}$ Ori C and
29.$^{\prime\prime}$2 from $\theta^{2}$ Ori A and it has the larger IF
observed in the Trapezium with a diameter of 2520 AU or
5.$^{\prime\prime}$6. One of the reasons could be because it is being
ionized by both stars as seen from the shape and orientation of its
cusp. Nevertheless, its disk is only 387 AU or 0.$^{\prime\prime}$86.
We can find enough cases of proplyds that lie in projection far from
$\theta^{1}$ Ori C and still have small IF, despite the large diameter
of their disks. For example, 072-135 is positioned at a distance of
176.$^{\prime\prime}$3 or 0.38 pc in the north-west direction from
$\theta^{1}$ Ori C and has a small IF of 1.$^{\prime\prime}$05 with a
large disk of 0.$^{\prime\prime}$75. At the same time, 176-543,
approximately at the same distance from $\theta^{1}$ Ori C as 244-440
or 141.$^{\prime\prime}$3, has an IF of only 1.$^{\prime\prime}$17 or
$\sim$ 1/5 of 244-440's chord diameter. Its disk is
0.$^{\prime\prime}$67 and R = 57\%.  In summary, there is no obvious
indication for larger proplyds to be located farther from the UV
source, even taking into consi\-de\-ra\-tion the unknown distant projection
correction.
In conclusion, it seems reasonable to adopt the average value of R,
$<$R$>$ = 49 $\pm$ 16\%, from our sample of 10 proplyds distributed at
different distances, and use it as a calibrator to compute the disks
diameters for the 125 bright proplyds without embedded silhouettes.
We find the same general result in the data from previous measurements by
O'Dell \cite{odell98} and Bally et al. \cite{bally98a}, discussed in the next section.

\subsection{Comparison with previous work}

In this section we make a detailed comparison of the data presented in
this paper and the observational data published in previous papers --
Bally et al. \cite{bally98a} , O'Dell \cite{odell98}, Johnstone et al. \cite{johnstone98}, St\"orzer
 \& Hollenbach \cite{storzer99} and Bally et al. \cite{bally00} (see Table 2). 

For the subsample of proplyds listed in these papers, and common to our
sample, an identical analysis was performed. Because the assumed
distances to the clusters used in the different papers are not always
the same (we use d$_{Orion}$ = 450 pc through out this paper) we
scaled all results to the same distance and only then compared them.
To compare the different subsamples of proplyds with embedded disks we
proceeded as follows: we computed the diameter ratio R for each
proplyd, the average diameter ratio $<$R$>$ for all the subsample of
proplyds, and the linear correlation coefficient \textit{r} between
the disk and IF diameters and the projected distances to the main
io\-ni\-zing stars\footnote{The correlation coefficient \textit{r}
is a measure of the linear association between variables. The strength
of this association is sometimes expressed as the square of the
correlation coefficient.  The resulting statistic is known as variance
explained. For example, a correlation of 0.5 means that (0.5)$^{2}$ or
25\% of the variance in Y is ``explained'' or predicted by the X
variable.  This parameter is used throughout this paper.}.

Bally et al. \cite{bally98a} observed four fields near the Trapezium with the 
Planetary Camera (PC), the 0.$^{\prime\prime}$05 angular resolution
portion of the WFPC2/HST, on March 1995. Observations were obtained
through narrow-band filters centered on the [S{\small II}], [N {\small II}],
H$_\alpha$, [O {\small I}] and [O {\small III}] lines and in an emission-line-free
continuum channel.
They present the dimensions of the various emission and opaque
components, and the peak surface brightness of the objects in each
emission line, for 43 Trapezium proplyds, of which 40 are bright (21
with dark disks embedded) and 3 are pure silhouettes. From these, 38
are common to our sample (35 bright proplyds + 3 pure silhouettes) and
19 are bright proplyds with dark embedded disks.
For the maximum projected distance to $\theta^{1}$ Ori C in the sample
(184-427 at 70$^{\prime\prime}$), the average diameter ratio $<$R$>$
computed for the 19 proplyds is 43 $\pm$ 18\%.  
Bally et al. \cite{bally98a} claims that, when both silhouette and
IF are seen, the ratio of the semi-major axis of the silhouettes
divided by the IF radius ranges from 0.25 to 0.67. This is also
confirmed in our analysis. The values of R show a large dispersion but
are in that range, with exceptions given to 158-327, 160-353 and
166-316.
The authors argue that there is a loose correlation between the cusp
radius and the projected distance to $\theta^{1}$ Ori C, with the cusp
radius increasing as \textit{r(d)$\varpropto$ d$^\alpha$} with
$\alpha$ = 0.5--0.8, where \textit{d} is the projected distance from
$\theta^{1}$ Ori C.  For the 34 bright proplyds common to our sample,
we find $\alpha$ = 0.46. The linear correlation coefficient between
the two variables is 0.54 or $\sim$ 30\%, meaning a loose correlation for
this specific sample of proplyds.

\begin{table}
\caption[c]{Orion proplyds observational parameters: comparison with previous papers.}
\label{table:2}
\tiny
\centering
\begin{tabular}{@{} lccccrccrr @{}}
\hline\hline
\noalign{\smallskip}
Paper & \multicolumn{2}{c}{Total} & \multicolumn{2}{c}{Common} &\multicolumn{2}{c}{max. d$_{proj}\theta^{1}$ OriC} & $<$R$>$ & \textit{r$_{IF}$}& \textit{r$_{disk}$}\\
\noalign{\smallskip}
& IF & disks & IF & disks & ($\prime\prime$) & (pc)& (\%) & (\%) & (\%)  \\
\noalign{\smallskip}
\hline
\noalign{\smallskip}
B98a$^{\mathrm{a}}$ & 40 & 21 & 34 & 19 & 70.0 & 0.15  (184-427) & 43$\pm$18 & 30 &  14\\ 
O98$^{\mathrm{b}}$ & 22 & -- & 20 & -- & 70.0 & 0.15 (184-427) & -- & 22 & -- \\
JHB98$^{\mathrm{c}}$ & 41 & 15 & 30 & 15 & 70.0 & 0.15  (184-427) & 48$\pm$22 & 42 & 7 \\
SH99$^{\mathrm{d}}$ & 10 & 10 & 10 & 10 & 56.7  & 0.12  (182-413) & 48$\pm$20 &  55 & 70\\
B00$^{\mathrm{e}}$ & 24 & 24 &18 & 18 & 176.3 & 0.38 (072-135) & 44$\pm$17 & 2 & 8 \\
\noalign{\smallskip}
\hline
\end{tabular}
\begin{list} {}{}
\item[$^{\mathrm{a}}$] Bally et al. 1998a
\item[$^{\mathrm{b}}$] O'Dell 1998
\item[$^{\mathrm{c}}$] Johnstone et al. 1998
\item[$^{\mathrm{d}}$] St\"orzer \& Hollenbach 1999 
\item[$^{\mathrm{e}}$] Bally et al. 2000
\item This table refers only to the sample of bright proplyds listed in previous papers and deliberately excludes the pure silhouettes. The second column refers to the total number of bright proplyds (here IF) and bright proplyds with embedded disks (here disks) listed in the papers. The third column refers to the subsample of objects that are common to our sample. $<$R$>$ is the diameter ratio computed for the common sample of embedded disks. The ma\-xi\-mum projected distances listed here are the ones measured in \textit{this} paper assuming 1 pix $\sim$ 0.$^{\prime\prime}$1 = 45 AU and d$_{Orion}$ = 450 pc; the differences between the different papers were negligible. The linear correlation coefficients \textit{r$_{IF}$} and \textit{r$_{disk}$} are represented by its square times 100\%. 
Bally et al. (2000) makes reference to many other proplyds but, we just consider the 24 H$_\alpha$ emission bright proplyds with dark disks embedded.
\end{list}
\end{table}
O'Dell \cite{odell98} presents a set of 22 bright proplyds observed in the
Bally's GO program 5469 with the HST/PC in 1995, of which 20 are
common to our sample and 19 to Bally et al. \cite{bally98a}. The author found
a very loose correlation of the proplyd cusp diameter with the
distance from $\theta^{1}$ Ori C, for a ma\-xi\-mum projected distance in
his sample of 70$^{\prime\prime}$ (184-427) or \-$\sim$ 0.15\- pc.  We
computed the linear correlation coefficient for the 20 common sources
and found a value of 0.47 or 22\%, indicator of a minor correlation
between the variables.

Johnstone et al. \cite{johnstone98} compare their model results with HST
observations from Bally et al. \cite{bally98a}. For 15 proplyds disks, for a
maximum projected distance to $\theta^{1}$ Ori C of
70$^{\prime\prime}$ (184-427), the average diameter ratio $<$R$>$
equals 48 $\pm$ 22\%. The correlation coefficient between disk
diameters and projected distance to $\theta^{1}$ Ori C is 0.26 or 7\%.
This paper presents a sample of 41 proplyds of which 30 (27 common to
our sample) were measured with the HST \textit{Planetary Camera} and
11 (3 common to our sample) with the \textit{Wide Field Camera}. The
total 30 bright proplyds IF diameters are correlated with the distance
to $\theta^{1}$ Ori C by 42\% or \textit{r} = 0.64, indicating a large
correlation between them.

St\"orzer \& Hollenbach \cite{storzer99} used the best 10 measured disks
reported by Johnstone et al. \cite{johnstone98} , in which both disk and IF
diameters could be observed (Bally et al. 1998a). For a maximum
projected distance of 56.$^{\prime\prime}$7 (182-413 or HST10),
$<$R$>$ = 48 $\pm$ 20\%. There is a very high correlation between both
disk and IF diameters and the projected distance to $\theta^{1}$ Ori
C, \textit{r$_{IF}$} = 55\% and \textit{r$_{disk}$} = 70\%. But, we
have to keep in mind that this is a very particular sample.

The last test was performed in the data from Bally et al. \cite{bally00}.
Although, the only diameters listed are for the 15 pure silhouettes,
they published images of all the embedded disks seen against the
bright ionization fronts: 24 disks in H$\alpha$, 16 disks in [O {\small III}]
and 2 disks in [N {\small II}]. The IF, and disk dia\-me\-ters of 17 proplyds
(disks boundaries are not clear in all pictures) were measured
directly from the published H$\alpha$ images. The proplyd 141-520
measures were taken from the [N {\small II}] i\-ma\-ge, increasing the sample,
from Bally et al. \cite{bally00} paper, to 18 bright proplyds.  There is no
significant correlation between the disk and IF dia\-me\-ters and the
projected distance to $\theta^{1}$ Ori C. For this sample:
\textit{r$_{IF}$} = 2\% and \textit{r$_{disk}$} = 8\%, for a maximum
projected distance of 176.$^{\prime\prime}$3 (072-135) or 0.38 pc.
The average diameter ratio is $<$R$>$ = 44 $\pm$ 17\% and the IF
diameters are, on average, 1.2 $\pm$ 0.2 times smaller than the ones
in this paper, in agreement with the 20\% increased resolution of
Bally et al. (2000) drizzled images.
For our sample of 135 bright proplyds, and separating them in
subsamples accordingly with the main ionizing O star, we computed the
linear correlation coefficient between the disk and IF diameters and
the projected distance to the main io\-ni\-zing source.  The first subsample is
composed of 105 bright proplyds which are being ionized primarily by
$\theta^{1}$ Ori C, and we found that there is no correlation between
the two variables for a maximum projected distance of
264$^{\prime\prime}$ (005-514) (Figure 1).  When we consider just the
proplyds which are at a distance less than 0.3 pc (St\"orzer \&
Hollenbach 1999), the correlation is low (-0.26 or 7\%).  For
the 19 proplyds ionized by $\theta^{2}$ Ori A, for a maximum projected
distance of 151$^{\prime\prime}$ (224-728), we get a week correlation
of 13\% for IF diameter with the projected distance to $\theta^{2}$
Ori A, even for distances less than 0.3 pc.  The disks show a higher
correlation of 18\%.
 
In summary, a dispersion of R for the Trapezium proplyds is found by
all the authors for the previous papers and, in all of them as well,
there is no obvious correlation between the IF ``chord diameter'' and
the projected distance to the Trapezium O stars.  Moreover, and within
the errors, all previous papers agree on the value of $<$R$>$.  It
seems reasonable to assume an average value for the diameter ratio
disk diameter/ IF diameter and use it as a ``calibrator'' to infer the
disks diameters for the 125 bright proplyds without visible disks.
The \textit{rms} in $<$R$>$ will be used to estimate the error in this
approximation.

\subsection{Caveats}
 
Selection effects make the detection of the disks difficult.  Dark
silhouettes can be easily overwhelmed by the unknown line-of-sight
contribution from the nebula or from young stars at their centers.
This is especially true for small ($<$ 2 WFPC2 pixels, or 90 AU)
disks.  On the other hand, small high surface brightness proplyds near
$\theta^{1}$ Ori C are hard to distinguish from point sources.  In
general, the location of a proplyd with res\-pect to the illuminating
stars strongly influences the visibility of embedded circumstellar
disks.  The procedure for defining disks and proplyds boundaries
assumed in this paper is systematic and precise but dependent on the
surrounding object nebular intensity, which is highly variable.
Nevertheless, this is still the best possible size measurement
procedure. Previous papers do not define the measuring procedure and
proplyds' sizes are even more dependent on personal judgment.
Estimating the completeness of our sample is a difficult task. Still,
given that the spatial resolution of the WFPC2 image is 67.5 AU, we
can crudely estimate that to a visual extinction limit of a few
magnitudes of A$_V$ our sample should be essentially complete for disk
diameters larger than $\sim$ 100--150 AU.

\label{sec:cluster_compare}


\begin{figure}[t]
\resizebox{10cm}{!}{\includegraphics[angle=0]{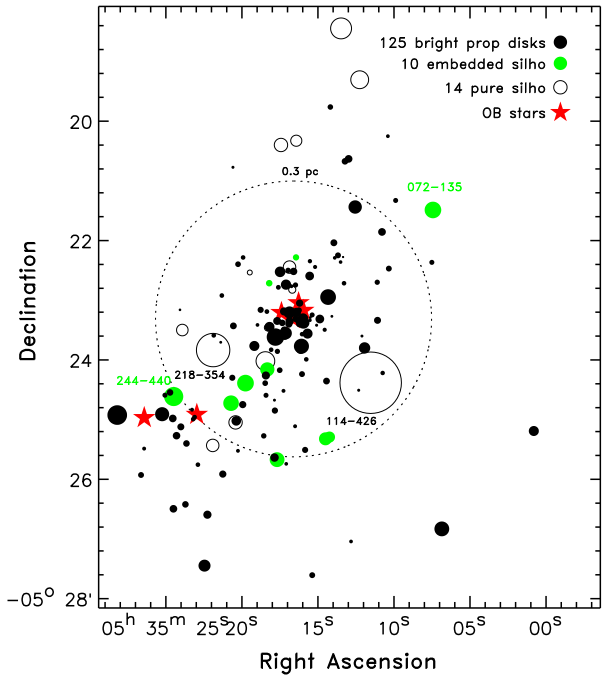}}
  \caption{Spatial distribution of total sample of 149 disks in the
     {Trapezium cluster}. The ``star'' symbols represent the 6 brightest
    OB stars. The filled circles indicate the positions of the 135
    bright proplyds' disks; gray (green) for the 10 proplyds with embedded
    silhouette disks and black for the remaining 125 from which the
    disk diameter is 49\% of the correspondant IF diameter.  The
    unfilled circles represent the positions of the 14 pure
    silhouettes.The diameters of the circles are proportional to the
    disk diameters they represent.} 
  \label{fig:spatial}
\end{figure}

%

\begin{figure}[t]
  \resizebox{9cm}{!}{\includegraphics[angle=0]{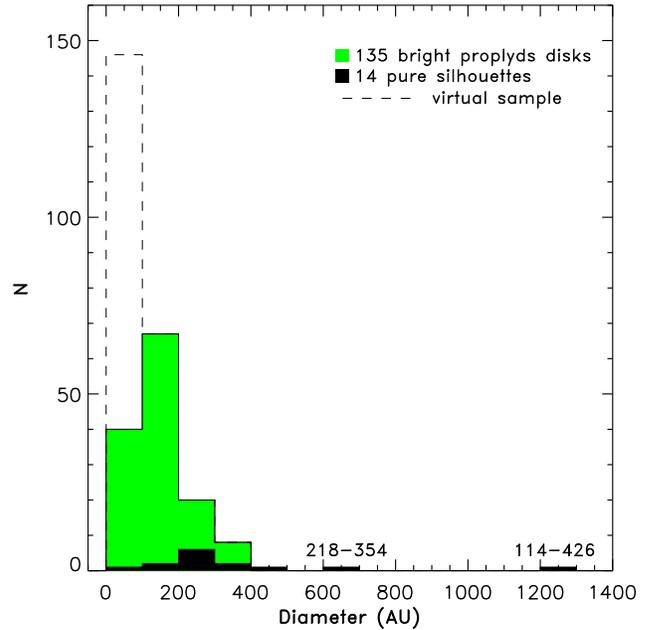}}
  \caption[]{Disk diameter distribution  for the total sample of 149 proplyds.
    The gray (green) filled histogram represents the 135 bright proplyds disks
    - 10 dark disks embedded in the bright proplyds' cusps and 125
    disks derived from the cusps diameters. The black filled histogram
    represents the 14 pure silhouettes seen only against the Orion
    Nebula background.  The first bin (0--100 AU) is highly incomplete
    due to the resolution limit of the data.  The virtual sample, in
    dashed line, was estimated based on the L-band excess of Trapezium
    sources. The large silhouet\-tes 218-354 and 114-426 are indicated.}
\label{fig:histo}
\end{figure}

%


\subsection{Size distribution of disks}

In Figure 4, we present the spatial diameter distribution of all the
 149 disks in our sample of the  Trapezium cluster.  The ``star''
symbols represent the 6 brightest Trapezium OB stars. The filled
circles indicate the positions of the 135 bright proplyds' disks;
gray for the 10 proplyds with embedded silhouette disks and black for
the remain 125 from which the disk diameter is 49\% of the
correspondent IF diameter. The unfilled circles represent the positions
of the 14 pure silhouettes. The diameters of the circles are
proportional to the disk diameters they represent. The pure
silhouettes are the largest disks that lie at relatively large
projected distances from the Trapezium. Two proplyds seem to fall off
the size distribution i) 114-426, with a diameter of 1242 AU, is by
far the  largest and ii) 218-354, with a diameter of 675 AU.   The
dashed circle indicates a radial distance of 0.3 pc from  $\theta^{1}$
Ori C and marks the limiting border of the FUV-dominated region of the
St\"orzer \& Hollenbach \cite{storzer99} photoevaporation model. It is also the
distance between the two ionizing stars, $\theta^{1}$ Ori C and
$\theta^{2}$ Ori A, with the EUV photo\-io\-ni\-zing luminosity of the
former 3-4 times greater than the later (O'Dell 2001b). About 73\% of
the proplyds in our sample are inside this region and 60\% at a
projected distance less than 0.2 pc.  The proplyd situated at the
largest distance from $\theta^{1}$ Ori C is 005-514 at $\sim$
264$^{\prime\prime}$ or 0.58 pc.  We do not find a correlation between
the disk diameters and the projected distance to $\theta^{1}$ Ori C,
as already discussed before.
\begin{figure}[t]
  \resizebox{9cm}{!}{\includegraphics[angle=0]{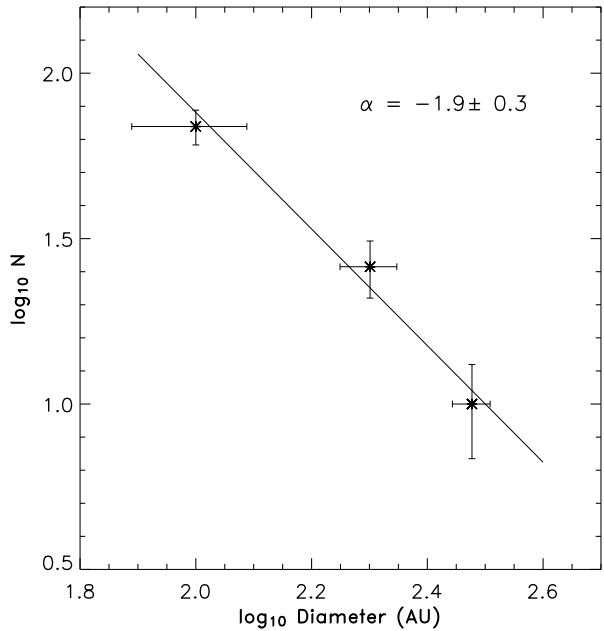}}
  \caption[]{Power-law fit to the size distribution of disks for diameters
    between 100--400 AU. The data can be described by a power-law with
    an exponent of $-$1.9 $\pm$ 0.3. }
  \label{fig:power}
\end{figure}

%
\begin{figure}[t]
  \resizebox{9cm}{!}{\includegraphics[angle=0]{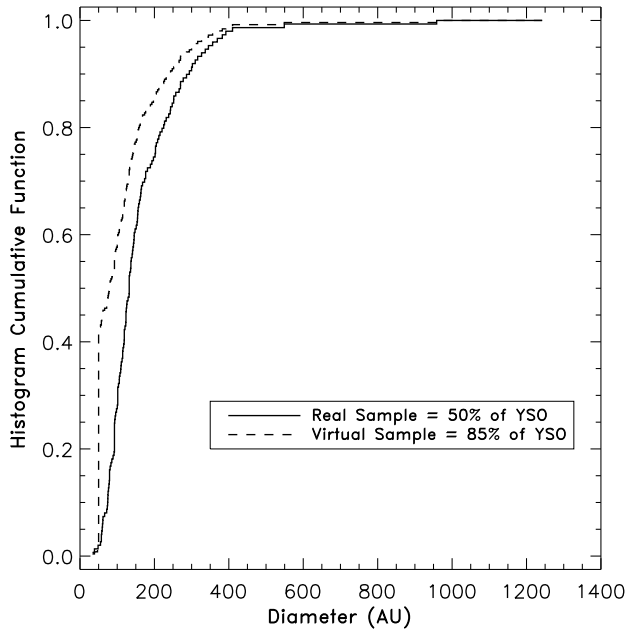}}
  \caption[]{Cumulative function of the histogram. The \textit{real} sample
    is composed by the 149 disks in this paper and represents
    $\sim$ 50\% of all Trapezium's YSO observed with the HST. The
    \textit{virtual} sample is the corrected sample accounting for the 35\%
    of HST unresolved disks and is composed by our sample corrected by
    adding 106 disks with diameters less than 50 AU.}
  \label{fig:cumulative}
\end{figure}

%
Figure 5 represents the disk diameter distribution histogram for the
total sample of 149 proplyds: 135 bright proplyds, in gray, and 14
silhouettes, in black.  To select a statistically significant binning
for the distribution we used a density kernel estimator (Silverman
1996) that indicated a bin-width of 100 AU at a 90\% confidence limit.
The resultant disk diameter distribution seems unimodal, with a tail
to the large diameters, suggesting a single population of disks that
is well characte\-ri\-zed by a power-law distribution.

Figure 6 represents the power-law fit, in a log-log scale, to the
sample of 149 disks presented in Figure 5. The first and incomplete
bin (0-100 AU) was deliberately excluded from the power-law fit.  We
assumed that all the disk diameters (ex\-clu\-ding silhouettes) are
49 $\pm$ 16\% of the proplyds' io\-ni\-za\-tion front diameters, based
on the 10 well-known cases with resolved embedded disks. The data is
well described by a power-law with an exponent of $-$1.9 $\pm$ 0.3.  To
determine the uncertainty on the power-law fit we performed
se\-ve\-ral tests, shifting the binning starting point by 25, 50 and
75 AU and the diameter ratio R for the 125 bright proplyds, to R =
33\% and 65\%, accoun\-ting for the dispersion found in this ratio.  The
power-law exponent is the average of the 12 coefficients determined in
these tests.  An obvious and unique deviation to the power-law
distribution characterizing our sample is the silhouette 114-426. With
a diameter of about 1242 AU this large disk seems to be an exception
compared to other proplyd disks in Orion.  About 80 to 90\% of
Trapezium's young stars and $\sim$ 65\% of the brown dwarfs show
infrared excess emission charac\-te\-ris\-tic of circumstellar disks
(Lada et al. 2000; Muench et al. 2001). Assuming that the fraction of
disks is not a strong function of depth into the Orion molecular cloud
(near-infrared surveys used by Lada et al. \cite{lada00} and Muench et al.
\cite{muench00} are less affected by dust extinction than the H$_\alpha$/HST
image analyzed in this paper) and given that the HST images reveal,
for about the same area of the cluster, that only $\sim$ 50\% of the
sources are associated with extended circumstellar structures, one can
infer that a re\-la\-ti\-vely large fraction of sources in the HST
images (30 to 40\%) also have disks but they are smaller than the
resolution limit of the HST image.  Since there were $\sim$ 300 YSO
imaged with the HST then $\sim$ 85\%, or about 255 of them, should have
circumstellar disks but we only identify 149. This means that the
other 106 disks are probably too small to have been resolved by the
HST.  The \textit{virtual} sample, or corrected sample of disks in the
Trapezium, is composed by the 149 disks from our sample plus 106 disks
of 50 AU added. The dashed line in Fig. 5\- represents the histogram for
the virtual sample of 255 circumstellar disks or 85\% of the YSO
observed in the  Trapezium cluster.  The cumulative disk diameter
function of the histogram in Fig. 5 is given by Fig. 7 and indicates
that 75 to 80\% of disks have diameters smaller than 150 AU (the
uncorrected value is about 60\%). This result is not consistent with
Rodmann \cite{rodmann02} who finds that 90\% of the disks have diameters smaller
than 80 AU (as referenced in Bate et al. 2003).  From this cumulative
function we can estimate that about 40\% of the disks in the Trapezium
have radius larger than 50 AU, the Solar System size.

\begin{figure}[t]
\resizebox{9cm}{!}{\includegraphics[angle=0]{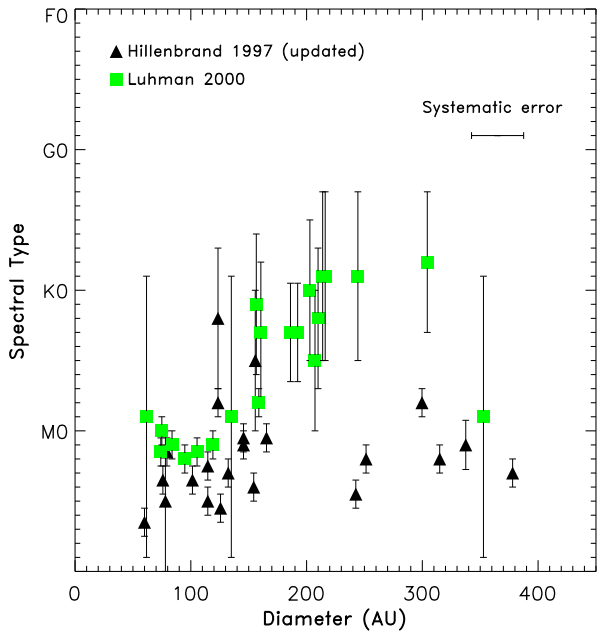}}
\caption[]{Disk diameter versus spectral-type of central parent star. 
  There is not an obvious correlation between the two variables.}
\label{fig:spectral}
\end{figure}

%


\subsection{Does disk size depend on the parent star mass?}
 
In Figure 8 we present disk diameter as a function of central star
spectral-type. Our proplyd list was compared with the Lynne
Hillenbrand's online working version of table1 (last update: September
2003) from Hillenbrand \cite{hillenbrand97}\footnote {\scriptsize
  http://www.astro.caltech.edu/$\sim$lah/papers/\rm$orion_{-}$main.table1.working}.
The selection criterion consisted in the overlap of the coordinates of
the proplys in the two tables with radius less than 2 arcsec.
For the resulting 52 sources in common, we considered the
spectral-types derived by Hillenbrand \cite{hillenbrand97} (optical spectroscopy)
and Luhman et al. \cite{luhman00} (NIR spectroscopy).  Only 43 of the 52
sources had a spectral classification and only 21 of them are
classified by Luhman et al. \cite{luhman00}.
For the 21 Luhman sources, represented in Figure 8 as gray squares, we
assumed the error to be equal to the uncertainty interval in Luhman's
NIR spectral classification.  The errors assumed for the 22
Hillenbrand sources are $\pm$1 spectral subclass, for spectral types
later than K7, and $\pm$1/2 class for spectral types former than K7,
as des\-cri\-bed in her paper. They are represented as black triangles in
the plot.  A systematic error of 45 AU or 1 pixel was considered in
the disk diameters.

In order to describe the strength of the linear asso\-cia\-tion between
the mass of the star and the diameter of the disk surrounding it,
several tests were performed on the data.  We calculated the Pearson
linear correlation coefficient for all the 43 sources of overlap
between our sample and Hillenbrand's ca\-ta\-log. To verify if this linear
correlation was or not dependent on the way that the YSOs spectral
types were determined, we separated the 43 combined data in two
samples: 21 proplyds from Luhman et al. \cite{luhman00}, and 22 proplyds from
Hillenbrand \cite{hillenbrand97}.  For the combined sample we derive a coefficient of
0.3 indicating a small correlation between the two variables (10\%).
However, we get a better correlation for Luhman sources alone
(suggesting a significant correlation between diameters and masses)
than in Hillenbrand sample where we get a very small correlation.
A larger sample or a better understanding of spectral ty\-ping differences
between optical and NIR should clarify this inconsistency.  We
conclude that, over the sampled mass range (late G to late M stars)
and with the present spectral data, there is not an obvious
correlation between disk diameter and stellar mass.

\section{Discussion}

An important result of this work is that there seems to exist one
single population of disks well characterized by a power-law
distribution.  Albeit the young age of the Trapezium, and given that
disk destruction is well underway, it is perhaps too late to tell if
the present day disk size distribution is primordial or if it is a
consequence of the massive star formation environment. Nevertheless,
the simple existence of a well defined disk size distribution hints at the
existence of a primordial disk size distribution. Intuitively one
would argue that a disk initial mass and size would be proportional to
stellar mass (continuum surveys performed in the mm/submm domain
suggest a tendency for the disk mass to increase with the stellar
mass, e.g., Natta et al. 2000) but the results expressed in Figure 8,
that there is not an obvious correlation between disk size and stellar
mass, are intriguing and suggest a more complex picture.  Most likely
there is a random disk destruction process, or a combination of processes,
that could be responsible for this lack of correlation.

The most important disks destruction processes are 1) viscous
accretion, 2) close stellar encounters, and 3) stellar winds.  Several
environment related disk destruction mechanisms (2 and 3) have
recently been proposed in the literature: Bate et al. \cite{bate03} performed
sophisticated hydrodynamical simulations of cluster formation and find
that most circumstellar disks are severely truncated by dynamical
encounters.  On the other hand, Richling \& Yorke \cite{richling00}, considering
photoevaporation, high initial disk masses, and fixed distances to
$\theta^{1}$ Ori C, obtain the present radius and masses observed in
Trapezium's proplyds, for a timescale consistent with the age of
$\theta^{1}$ Ori C ($\sim$ 0.5 Myr).  Also, Scally \& Clarke \cite{scally01}
performed numerical N-body simulations of the  Orion Nebula cluster and
determine a very low probability of an encounter at a d $<$ 100 AU, at
the cluster's present age, suggesting photoevaporation as the most
significant disk destruction mechanism. If significant UV radiation is
available, photoevaporation is the dominant disk dispersal mechanism
(Hollenbach et al. 2000). The intense UV radiation field heats the
disk surface, drives mass-loss and produces the bright ionization
fronts, on a relatively short timescale.
For e\-xam\-ple, the Trapezium proplyds have radial velocities of 24 to 30
Kms$^{-1}$ (Henney \& O'Dell 1999) and inferred disk masses of
0.005-0.02 M$_\odot$ (Lada et al. 1996; Bally et al. 1998b), with  the 0.02
M$_\odot$ upper bound determined for the largest pure silhouette
114-426.  Mass-loss rates of order 10$^{-7}$ to 10$^{-6}$
M$_\odot$yr$^{-1}$ have been measured for a number of proplyds by
several groups (Churchwell et al. 1987; Johnstone et al. 1997; Bally
et al. 1998b; Johnstone et al. 1998; Henney \& Arthur 1998; Henney \&
O'Dell 1999) with a mean mass-loss rate of 3.3 $\times$ 10$^{-7}$
M$_\odot$yr$^{-1}$ for all 31 proplyds that have been studied with
sufficient detail (out of a total of $\sim$ 150 Orion bright
proplyds).  This implies that a minimum Solar Nebula mass disk (0.01
M$_\odot$) will loose half of its mass in 10$^{4}$ to 10$^{5}$ years
and even less for the less massive observed objects. Such lifetimes
are quite small compared to the estimated age of $\theta^{1}$ Ori C or
0.5 Myr. Herbig \& Terndrup \cite{herbig86} and Hillenbrand \cite{hillenbrand97} find that
the ages for low-mass stars in the  Orion Nebula cluster range from 3 x
10$^5$ to 10$^6$ yr, with only a few as young as 10$^5$ yr.  This
brings a major conundrum, since proplyds exist and have been surviving
the photoevaporation.  There are two possible explanations: either
$\theta^{1}$ Ori C is very young,
or the illumination time of the proplyds is short (which would require
a large spatial motion of $\theta^{1}$ Ori C). Recently, Tan \cite{tan04}
claimed that the BN object is a runaway B star dynamical ejection
event by the $\theta^{1}$ Ori C star, just $\sim$ 4000 years
ago. Regardless, the point is that photoevaporation is likely to be
the dominant disk destruction process.  Still, relevant observational
studies of the dynamics of embedded stellar clusters are not available
yet (efficient near-infrared high-resolution spectrographs are only
now becoming available) and it is too soon to dismiss close stellar
encounters as an important disk destruction process in young clusters.

\subsection{The Trapezium as the birthplace of Solar system analogues}

\textit {Can planet formation endure disk destruction mechanisms?}

If the growth of larger particles can occur before the removal of gas and
small particles, planets may nevertheless form inside the dust disks
embedded in such an adverse environment.  There are evidences for
grain growth in the largest silhouette 114-426 (Throop et al. 2001;
Shuping et al. 2003).  Recently, Throop \& Bally \cite{throop05} proposed that
UV radiation can stimulate the rapid formation of planetesimals in
externally-illuminated protoplanetary disks. It might then be that
photoionization has an overall positive feedback on planet formation,
while being an important (gas) disk destruction process. If this is
the case, this would have far reaching implications to planet
formation in the Galaxy, and likely in the Universe, since most stars
are born in clusters containing O and B stars (Lada \& Lada 2003). In
the specific case of the Trapezium, where about half of the disks are
larger or about the Solar system size (see Section 3.7), one could
expect pla\-ne\-tary systems to be common and not fundamentally different
from the Solar system itself.

It is striking that the remnant Solar system disk appears to be
truncated, not unlike the Trapezium disk discussed in this paper. It
is now a rather well established and intriguing fact that the Kuiper
belt, which is a repository of the solar system's most primitive
matter, has a well defined outer edge at about 50 AU (e.g., Jewitt 
2002). At least three mechanisms for its origin have been
proposed, none of which has raised the ge\-ne\-ral consensus of the
community of the experts (see Morbidelli et al.  2003 for a review):
1) a dynamical origin involving a 1 Gyr lived Mars mass at object at
60 AU (Brunini \& Melita 2002), 2) di\-f\-fe\-ren\-ti\-al accretion rates
producing a disk edge (Weidenschilling 2003), and 3) planetesimal
disk truncation by the passage of a star in the vicinity of the Sun
(e.g., Ida et al. 2000, Kobayashi \& Ida 2001). The HST images of
the  Trapezium cluster disks, in particular the silhouette disks,
present us with clear e\-vi\-den\-ce that these disks seem to have already
well defined edges at their relatively very young ages. If the Sun was
born in a stellar cluster, the  probable case as most stars are born in
clusters, this is suggestive that the origin of the Kuiper belt outer
edge is likely to be due to the star formation environment and disk
destruction processes (photoevaporation, collisions) present in
archetypal star formation factories such as the Trapezium. If this is
the case, these well defined outer edges are imprinted in these disks
at the earliest phases of their evolution, even before the formation
of Kuiper belt-like objects.

The results in this paper, together with future higher-resolution
observations and modeling, can help identify the dominant process for
circumstellar disk destruction and provide insights into the survival
rate of circumstellar material surrounding the YSOs, and therefore,
insights on star and planet formation in very young clusters
containing O and B stars, the typical nursery for most stars in
the Galaxy.

\section{Conclusions}

The main results in this paper can be summarized as follow:

\begin{itemize}
\item There is no meaningful correlation between the IF ``chord
  diameters'' and the projected distance to $\theta^{1}$ Ori C.  We
  find the same tendency in our analysis of previously published data.

\item Direct measurements of 10 disks embedded in bright proplyds show
  a great dispersion in the diameter ratio R = disk diameter/ IF
  diameter, and R seems not to be correlated with the projected
  distance to $\theta^{1}$ Ori C. 244-440 (R = 15\%) and 072-135 (R =
  71\%) illustrate the 2 opposite ``extreme'' cases.

\item Assuming R = 49 $\pm$ 16\% (the average ratio from the 10 cases)
  to compute the disk sizes for the 125 bright proplyds we determined
  an unimodal disk size distribution, representing a single population
  of disks, that is well charac\-te\-ri\-zed by a power-law distribution
  with exponent of -1.9 $\pm$ 0.3.

\item For the stellar mass sampled (from late G to late M stars) we
  find that there is no obvious correlation between disk size and
  stellar mass.

\item The pure silhouettes are clearly the largest disks and have
  large projected distances to $\theta^{1}$ Ori C. In particular,
  object 114-426 is rather unique given its size and it falls off the
  above characterized disk size distribution.

\item We estimate that about 40\% to 45\% of the Trapezium cluster
  disks have radius larger than 50 AU, the Solar System size.

\item We suggest that the origin of the Solar system's (Kuiper belt)
  outer edge is likely to be due to the star formation environment and
  disk destruction processes (photoevaporation, collisions) present in
  the stellar cluster on which the Sun was probably formed.  

\item We identified a previously unknown proplyd and
  named it 266-557, following convention.

\end{itemize}

This statistical analysis should be repeated in function of "true"
distances to the OB stars and not the projected ones.  Proplyd
geometry and true distances will lead to the 3-D spatial distribution
of proplyds in Orion.

\begin{acknowledgements} The authors are grateful to  Nicole Homeier,
  Michael Liu, Martino Romaniello, Herve Bouy and Ricardo Demarco for fruitful discussions.
\end{acknowledgements}
\end{document}